\documentclass[sigconf]{acmart}
\AtBeginDocument{%
  }

\acmConference{}

\usepackage{multirow} 
\usepackage{color}
\definecolor{ecolor}{rgb}     {0.0,0.0,1.0} 

\definecolor{ycolor}{rgb}     {0.0,0.0,0.0} 
\newcommand\yc[1]{\textcolor{ycolor}{#1}}

\usepackage{xcolor}
\usepackage{graphicx}
\usepackage{subcaption}
\usepackage{booktabs}
\usepackage{tabularx}
\usepackage{makecell}

\begin{document}

\title{/UnmuteAll: Modeling Verbal Communication Patterns of Collaborative Contexts in MOBA Games}

\author{Yongchan Son}
\affiliation{
\institution{Chung-Ang University}
  \city{Seoul}
  \country{Republic of Korea}
}
\email{daniel0801@cau.ac.kr}
\author{Jahun Jang}
\affiliation{
\institution{Chung-Ang University}
  \city{Seoul}
  \country{Republic of Korea}
}
\email{wkdwkgns213@cau.ac.kr}
\author{Been An}
\affiliation{
\institution{Chung-Ang University}
  \city{Seoul}
  \country{Republic of Korea}
}
\email{dksqls08@cau.ac.kr}
\author{Jimoon Kang}
\affiliation{
\institution{Korea University}
  \city{Seoul}
  \country{Republic of Korea}
}
\email{kangjimoon@korea.ac.kr}
\author{Eunji Park}
\authornote{Corresponding author}

\affiliation{
\institution{Chung-Ang University}
  \city{Seoul}
  \country{Republic of Korea}
}
\email{eunjipark@cau.ac.kr}
\renewcommand{\shortauthors}{Son et al.}


\begin{abstract}
Team communication plays a vital role in supporting collaboration in multiplayer online games. Therefore, numerous studies were conducted to examine communication patterns in esports teams. While non-verbal communication has been extensively investigated, research on assessing voice-based verbal communication patterns remains relatively understudied. In this study, we propose a framework that automatically assesses verbal communication patterns by constructing networks with utterances transcribed from voice recordings. Through a data collection study, we obtained 84 game sessions from five \textit{League of Legends} teams and subsequently investigated how verbal communication patterns varied across different conditions. As a result, we revealed that esports players exhibited broader and more balanced participation in collaborative situations, increased utterances over time with the largest rise in decision making, and team-level differences that were contingent on effective professional training. Building upon these findings, this study provides a generalizable tool for analyzing effective team communication.
\end{abstract}

\maketitle

\section{Introduction}
Multiplayer Online Battle Arena (MOBA) is a real-time strategy game genre in which each player selects a character, levels up, and strengthens their character to destroy the opponent's base. 
Therefore, communication, such as sharing information and issuing directives, is crucial for executing team-level strategies in MOBA games. 
In particular, when a team engages in battles against the opponent team or must coordinate to accomplish difficult tasks within a short period of time (e.g., killing powerful neutral monsters), more intensive collaboration is required. 
Since MOBA games progress in real time, players have to quickly exchange information and make rapid collective decisions to adapt to changing game situations. 
Without timely and effective communication, teams are more likely to miscoordinate, lose critical objectives, and ultimately reduce their chances of winning. 
Accordingly, professional esports teams provide coaching to help players communicate efficiently and effectively~\cite{chang2024systematic, castro2021didactic}.

In the field of Human-Computer Interaction (HCI), there has been active research on analyzing communication patterns in MOBA games.
However, most previous works have primarily focused on non-verbal communication such as pings and emote~\cite{leavitt2016ping,zheng2023use}.
While non-verbal communication can serve as an efficient channel during gameplay, its limited semantic richness constrains the expression of nuanced intentions which may lead to misunderstandings~\cite{lee2025less}.
Therefore, players prefer voice chat for cooperation due to its fast, natural, multitasking-friendly nature~\cite{vaddi2016investigating}.
Aligned with this perspective, recent studies have explored the relationship between verbal communication patterns (e.g., frequency of specific utterances) and players’ perceptions of team cohesion~\cite{tan2022communication, maier2024talking}.
However, studies analyzing voice-based communication in the context of team collaboration have primarily relied on qualitative methods, such as self-reports and observational interviews. 
Accordingly, research that automatically analyzes voice communication patterns during gameplay remains limited.
Drawing on this motivation, this study addresses the following research questions:

\begin{itemize}
    \item \textbf{RQ1}: How can we model verbal communication patterns based on the utterances exchanged by esports players during gameplay?
    \item \textbf{RQ2-a}: How do verbal communication patterns vary across different game contexts (e.g., situations requiring collaboration, temporal phases of gameplay)?
    \item \textbf{RQ2-b}: How do verbal communication patterns differ between teams with and without professional communication training?
\end{itemize}

To answer the first research question, we proposed a framework that quantitatively analyzes players' verbal communication based on in-game and audio data. 
First, we utilized in-game data to identify salient situations requiring cooperation among multiple players, referred to as Moments of Interest (MoI). 
In addition, using audio data containing players' recorded voices, we transcribed speech, segmented utterances, and classified each utterance according to its communicative intent (i.e., dialogue act classification). 
Based on the classified utterances, we then structured the conversation between players within a specific time window into communication networks. 
To address the second research question, we calculated widely used quantitative metrics (i.e., density and centralization) from the modeled networks to infer team-level communication characteristics. 
Using these metrics, we first examined whether communication patterns differ between situations that require collaboration and those that do not (i.e., MoI vs. non-MoI) and how these patterns changed over the progression of the match. 
We further analyzed differences between professional teams that had received communication training and amateur teams that had not, and compared communication patterns across teams.

To collect data, we implemented a lightweight logging system that captures both in-game and audio communication data with synchronized timestamps. 
To capture in-game data on the tournament servers used by professional players, we utilize the Live Client Data API~\cite{Riot2025}. 
In addition, to record audio in the environment most familiar to the players, we developed a bot in Discord along with an additional timestamping module to collect players’ voice communication data.
The logging system was designed to impose minimal overhead on the Operating System (OS) to ensure that gameplay was not affected.
For the data collection study, we recruited a total of 25 participants, including 15 professional \textit{League of Legends} (LoL) players from three teams and 10 amateur LoL players from two teams. 
We recruited participants at the team level and selected teams that had already been formed and had been playing prior to the data collection study. 
Furthermore, to minimize communication differences due to game skill disparities, we recruited teams of similar ranks. As a result, the average rank of the professional teams was Grandmaster (top 0.051\%), while that of the amateur teams was Master (top 0.5\%).
A total of 84 games were collected and utilized for the analysis, comprising 41 games from professional team A, 11 games from professional team B, 13 games from professional team C, 10 games from amateur team A, and 9 games from amateur team B.

As a result, we revealed that verbal communication patterns varied across the three conditions.
First, in situations requiring collaboration, more players participated in conversation, and utterances were distributed more evenly compared to situations not requiring collaboration.
In addition, as the game progressed, the frequency of utterances increased, with utterances associated with action negotiation showing the most sharpest rise.
Moreover, professional teams showed a tendency for directive patterns to increase and question-based information gathering patterns to decrease as the game progressed.
In professional teams, more players were involved in communication related to information sharing and decision-making, and their contributions were more evenly distributed.
However, the manifestation of this tendency was contingent on training effectiveness, and professional team B did not display these patterns.
The analysis results can be applied to diagnose the state of team communication, support player selection, and inform training practices. Furthermore, the proposed framework has broader applicability as a generalizable tool for analyzing Computer-Mediated Communication (CMC).

In this study, we make the following contributions: 
\begin{itemize}
    \item We propose a framework that automatically assesses verbal communication patterns utilizing in-game and voice audio data.
    \item We provide novel insights into how team communication adapts to collaborative demands and game contexts.
\end{itemize}

\section{Related Works}
\subsection{Computer-Mediated Communication in Esports}
Esports has gained increasing attention in the HCI community as a novel testbed for \textit{virtual teams} communication research.
This is because the esports team is a representative example of \textit{virtual teams} that can form and collaborate without physical co-location.
In general, \textit{virtual teams} primarily communicate through networked systems (i.e., CMC~\cite{romiszowski2013computer}).
Similarly, in esports, players utilize various forms of CMC (e.g., verbal and non-verbal communication) to share information about ongoing in-game situations.
Consequently, the use of CMC makes it feasible to collect data that reflects the interplay between in-game events and team communication.~\cite{block2018narrative, wadley2015voice}.
Thus, prior HCI studies have examined CMC in esports contexts.
Among these, pings have been the most widely studied form of non-verbal communication.
For example, Leavitt et al.~\cite{leavitt2016ping} revealed a concave relationship between pings and performance; increased frequency of pings enhances performance up to a point before becoming detrimental.
Another study differentiated types of pings and identified which ones positively or negatively influence team performance~\cite{zheng2023use}.
Moreover, Wuertz et al.~\cite{wuertz2017players} conducted a survey to understand how and why players use pings.
Non-verbal communication, however, is often ambiguous, as pings can be misinterpreted and fail to keep pace with rapidly changing game contexts~\cite{lee2025less}.

\yc{Verbal communication in games has also been the focus of prior research.
Continuous and active conversation between players during gameplay constitutes verbal communication in esports, comprising both text chat and voice chat. 
Prior research has examined the meanings embedded in text chat.
For example, Kim et al.~\cite{kim2024communication} found that resilience and toxicity measured from text chat had positive and negative impacts on team performance, respectively.
Lee et al.~\cite{lee2025less} also observed in-game text chat to understand players' perception of communication attempts.
Voice communication such as voice chat enables the efficient and effective transfer of large amounts of information in rapidly changing situations and serves as an important means of fostering social connectedness among players~\cite{wadley2015voice}.
In fact, players prefer voice communication to text chat while performing collaborative tasks during gameplay because of its ease, speed, naturalness, and compatibility with multitasking~\cite{vaddi2016investigating}. In line with this perspective, recent studies have examined the relationship between voice communication patterns and players' perceptions of team cohesion~\cite{tan2022communication, maier2024talking}. However, these studies primarily relied on qualitative methods, such as self-reports~\cite{maier2024talking} or exploratory qualitative case studies~\cite{tan2022communication, kostrna2025textual}. Accordingly, research that automatically assesses voice communication patterns during gameplay is lacking.}

\yc{In this study, we examine whether verbal communication patterns assessed from voice recordings align with the findings from prior qualitative research. Prior observational study indicated that esports players are more likely to rely on text chat during collaborative moments to discuss more complicated tactics~\cite{lee2025less}. They also indicated that such collaborative moments occur more frequently as the game advances into its later phase. Moreover, esports coaches actively provide feedback to ensure that players communicate both efficiently and effectively~\cite{tan2022communication, maier2024talking}. Given these contexts, we formulate three hypotheses:}
\begin{itemize}
    \item \textbf{H1}: Verbal communication patterns will differ in situations requiring collaboration.
    \item \textbf{H2}: Verbal communication patterns will vary across temporal phases of gameplay.
    \item \textbf{H3}: Professionally trained teams will exhibit different verbal communication patterns.
\end{itemize}

\subsection{Communication Network}
Communication networks refer to the patterns of contact created by the flow of information between communicators~\cite{monge2003theories}, and have been widely employed to model interactions that occur in CMC.~\cite{rice1990computer}.
Social Network Analysis (SNA) serves as a tool for examining such networks by representing people as nodes and their interactions as edges, which enables the systematic study of communication structures~\cite{tabassum2018social}. 
SNA is particularly useful in providing insights at both the individual and network levels through the use of statistical metrics. 

Network-level metrics such as density and centralization serve as an effective measure for team-level analysis as they provide assessments of the overall structure of the network. 
Density is measured via the ratio of existing connections (i.e.,  edges) between individuals to the possible connections of the network, and reflects the social level of team cohesion~\cite{newman2003structure}. The density–performance hypothesis~\cite{balkundi2006ties} suggests that teams with higher network density reflect more frequent and reciprocal interactions among members, and therefore tend to perform better. Dense networks are associated with increased interdependence~\cite{sparrowe2001social}, enhanced coordination and cooperation ~\cite{molm1994dependence}, and greater trust and information sharing~\cite{littlepage1997effects}, all of which contribute to improved team performance. 

Network centralization is defined as the degree to which communication is concentrated in one or a few members rather than being evenly distributed across all members of the organization~\cite{freeman1978centrality}. Findings on the impact of centralization vary across studies, with no clear consensus in the literature. Previous studies indicated that as information flows primarily through central actors in a centralized network, their biases may propagate across the team and increase the risk of flawed collective judgments ~\cite{becker2017network}. 
Moreover, decentralized structures possess numerous communication pathways among team members, which is beneficial for adapting to faster information flows~\cite{bernstein2016beyond, lee2017self}. However, a recent experimental study using a cooperative video game demonstrated that centralized network structures enhance collective adaptability in a shifting environment, where optimal strategies for success change dynamically over time ~\cite{bernstein2023network}. Central nodes in such structures maintain the relative independence of peripheral nodes and mitigate conformity pressure, which fosters the generation of diverse solutions and the effective dissemination of ideas~\cite{bernstein2023network}.
Density and centralization are complementary to each other, as density explains the connectedness of a network, and centralization represents the concentration of connectedness around a particular node~\cite{wasserman1994social}.
Node-level metrics (e.g., betweenness, closeness) can also be adopted to analyze the position of a vertex and therefore identify key players in the network~\cite{freeman1978centrality}. 

SNA has been widely applied across domains characterized by intra-group social interactions, including business settings~\cite{bonchi2011social}, team management~\cite{rosenthal1997social}, and online social platforms such as YouTube or Instagram~\cite{wattenhofer2012youtube, vassey2023cigarette}. Among these, team sports have garnered particular research interest due to the direct link between collaboration and performance~\cite{yamamoto2011common, cintia2015network}. In traditional sports like soccer and basketball, prior work has modeled player interactions, such as passes or assists, as edges within networks to predict team performance using network metrics~\cite{grund2012network, cintia2015network, rein2017pass}. As interest in traditional sports grew, recent studies have extended SNA to its computer-mediated counterpart, esports.
For example, kill–assist relationships in LoL have been used to construct interaction networks, allowing researchers to investigate the connection between network structure and team efficiency or performance \cite{mora2019team}. Other work has applied SNA to explore psychological dimensions (e.g., coping strategies, burnout) in professional esports contexts \cite{poulus2024burnout}. Furthermore, SNA can be used to model verbal communicative interactions within teams, as shown in studies linking verbal dynamics with social influence or behavioral patterns \cite{bekiari2017insights, bekiari2015verbal}. Building on this foundation, the present study constructs communication networks based on voice communication occurring under specific in-game conditions and investigates verbal communication patterns via statistical metrics.
\section{Game play and communication logging system}
\begin{figure*}[t]
\centering
  \includegraphics[width=1\linewidth]{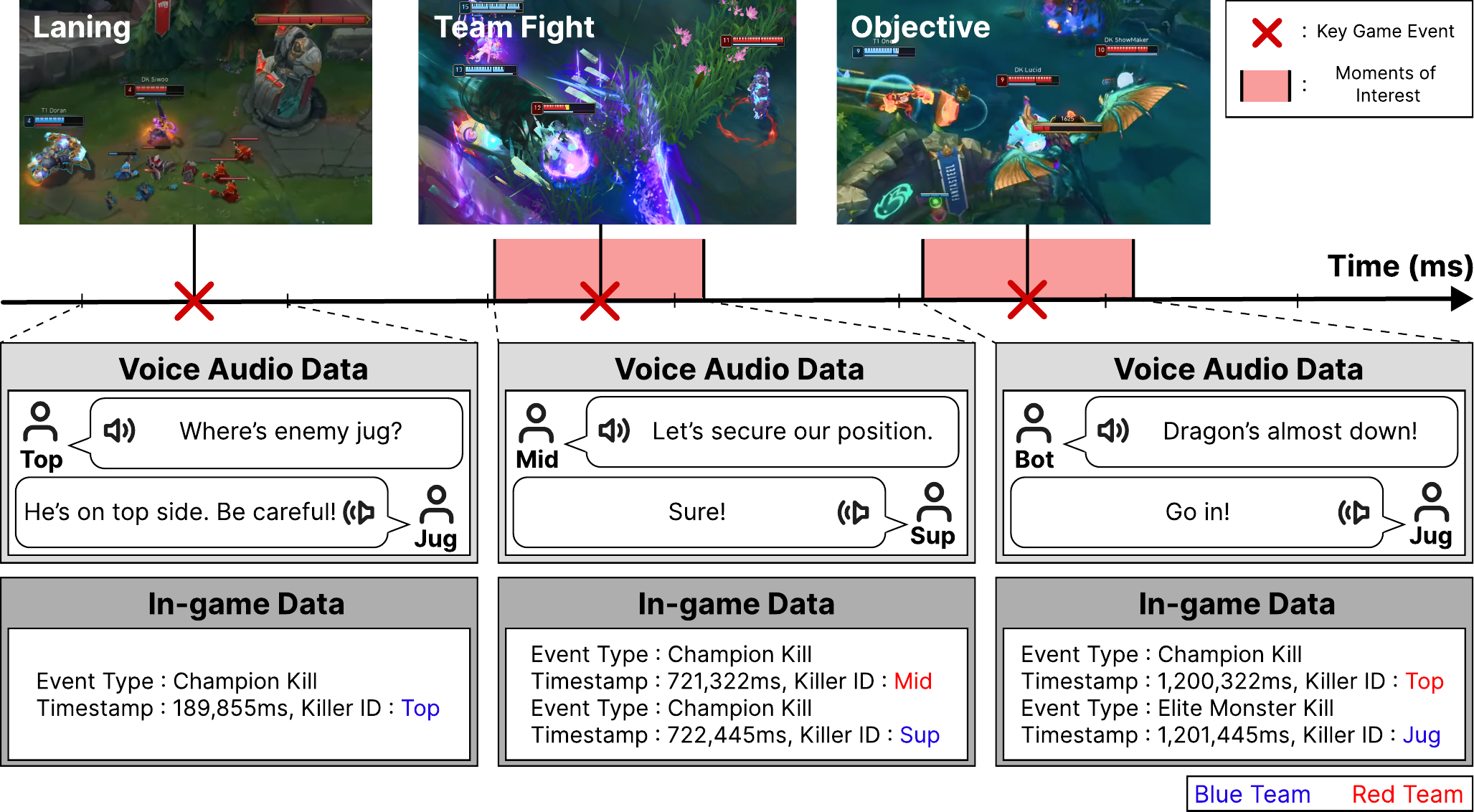}
  \caption{Overview of the collected data from the logging system: Voice audio data and in-game data are collected using the logger system. In-game data includes players' information, in-game events, and game time.}
  \label{fig:logger}
  \Description{Figure 1: Figure 1 illustrates the collected data from the logging system. Our logging system collects in-game data and audio data. In-game data includes players' information, in-game events such as Champion Kills or Elite Monster Kill that happen during laning, team fights, and objective contests. The role of the player and the transcripts are synchronized with each in-game event.}
\end{figure*}
We developed a logging system to collect data on various in-game events and communication among players. 
To synchronize in-game and audio data, we also logged timestamps concurrently. Detailed descriptions of each data acquisition method are provided below. 
\subsection{In-Game Data}
\subsubsection{\textbf{Logging System}}
We developed an in-game data logging system using the Live Client Data API~\cite{Riot2025}. 
The Live Client Data API exposes game state information (e.g., in-game events, players' stats) through a local server while the client is running, and our logging system periodically polls this data.
Designed as a lightweight background process, the logging system retrieves information without affecting gameplay. 
Although the Riot API has been widely used in previous studies~\cite{lee2024characterizing, cordova2024predicting}, it does not support in-game data collection on the tournament servers used by professional players. 
As our study involves professional players, we employed the Live Client Data API, which supports data collection on tournament servers.

When a game starts, the logging system detects the first valid response from the LoL client's API request and initiates data collection. Following the initiation, in-game events and timestamps were retrieved at one-second intervals. The game start time is recorded in Universal Time Coordinated (UTC), while the in-game duration is logged using local timestamps (i.e., the game start is set to 0 ms). Both measures were recorded with millisecond precision. A distinctive feature of the custom games, unlike regular ranked games, is the availability of a pause function. For example, coaches or referees may invoke this function to temporarily halt the game in cases such as hardware malfunctions, network issues, or in-game errors. We logged both the time at which a pause is initiated and released in UTC. The logging system remains active until the LoL client is terminated. Upon termination of each game, the system enters standby mode to await the signal of the next game. When the game ends, all logged data from completed games is automatically packaged and sent to a preregistered email. As it is crucial for professional players that the system does not introduce any delays during gameplay, we design the logging system to impose only a minimal operational burden on the OS and verified its latency performance. In practice, the system exhibited low average resource usage: CPU utilization (0.18\%), resident set size (33.34 MB), and virtual memory size (573.24 MB).

\subsubsection{\textbf{Details of Collected Data Items}}
The logger saves (1) players' information, (2) in-game events, and (3) game time in CSV format.
First, players' information contains username and numerical identifiers (i.e., 1–5 for the blue team, 6–10 for the red team) that allow identifying each player's role. The identifiers are assigned in standard position order (i.e., Top, Jungle, Mid, Bot, and Support).
For the in-game events, we logged event type (e.g., \texttt{Champion Kill}, \texttt{Elite Monster Kill}, or \texttt{Building Destruction}), numerical identifiers of the players who are involved in each event, and the time an event occurred. 
Specifically, for \texttt{Champion Kill}, the identifiers of the killer (i.e., the player who delivered the final blow), the victim, and any assisting players who contributed damage or provided supportive effects (e.g., buffs) are recorded. 
For \texttt{Elite Monster Kill} and \texttt{Building Destruction}, the identifiers of the killer and assisting players are stored.
The occurrence time of each event is recorded as a local timestamp. 
Lastly, the game time log stores the start time of the match in UTC, the total duration of the game, and the start and end times of pauses that occurred during gameplay.

\subsection{Audio Data}
Most LoL users communicate with their teammates via Discord~\cite{discord}, and almost all professional players rely on Discord as their primary communication channel~\cite{lipovaya2018coordination}. 
Therefore, we recorded voice communication during gameplay using the Craig Bot~\cite{Craig} on Discord. 
Craig is a recording bot on Discord that supports multi-track recording (i.e., five separate audio files).
When five players join the same channel with the Craig Bot, the bot records each player's voice as audio data. However, since the Craig Bot does not log timestamps, synchronizing audio with in-game data is difficult when using the Craig Bot alone. To address this, we developed an additional bot (i.e., Checker Bot) that records the timestamp in UTC when the Craig Bot begins recording. 
As a result, at the end of each match, five individual audio files in Adaptive Differential Pulse Code Modulation (ADPCM)~\cite{cummiskey1973adaptive} WAV format are compressed into a ZIP file and automatically uploaded to Google Drive, with the text file, which contains a timestamp marking the start of the recording.
Details on the data collected from the logging system are illustrated in Fig. \ref{fig:logger}.

\begin{figure*}[t]
\centering
  \includegraphics[width=1\linewidth]{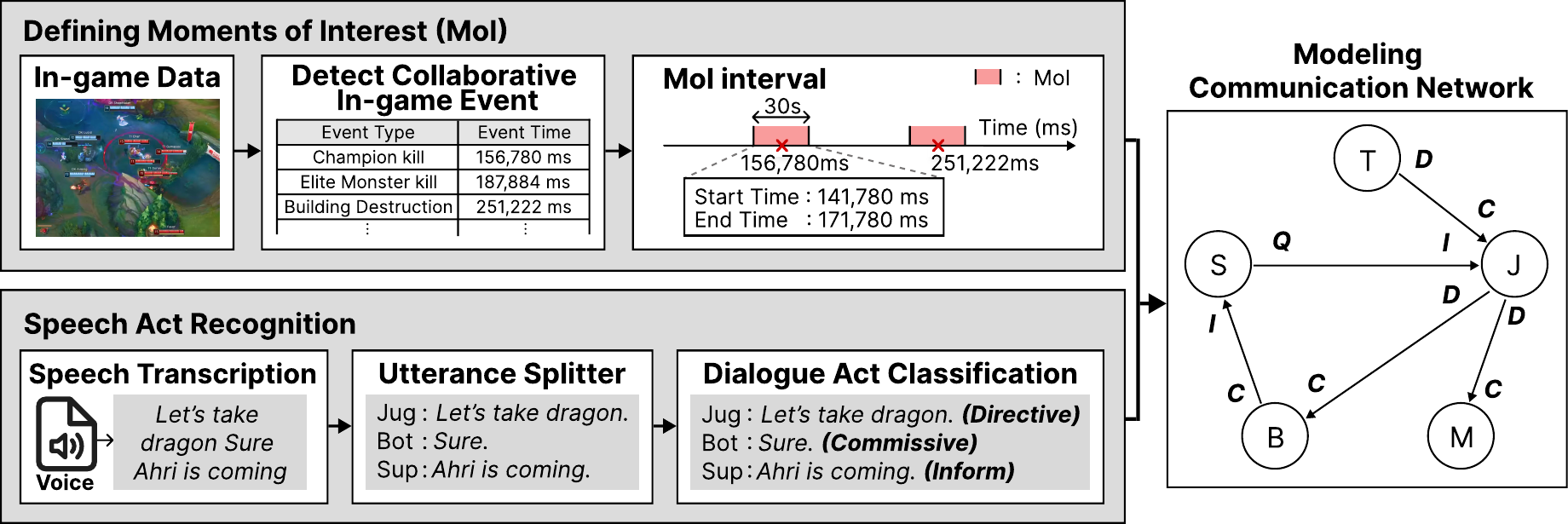}
  \caption{Overview of the framework: Collaborative in-game events (e.g., \texttt{Champion Kill}, \texttt{Elite Monster Kill}, \texttt{Building Destruction}) are defined as MoI with 30-second intervals spanning 15 seconds before and after each event. Audio data are transcribed using a speech transcription pipeline, and segmented into utterances, which are then classified into one of four Dialogue Acts (DAs): Inform (\textbf{I}), Question (\textbf{Q}), Directive (\textbf{D}), or Commissive (\textbf{C}). Within each MoI interval, utterances and their DAs are ordered by start time, and adjacency pairs are modeled as edges while players serve as nodes in the resulting communication networks. Player roles are denoted as T (Top), J (Jungle), M (Mid), B (Bot), and S (Support).}
  \label{fig:framework}
  \Description{Figure 2: Figure 2 is an overview of the framework in this study. The framework comprises three components. First, collaborative in-game events are detected from in-game data, which are used to determine the MoI interval, which is 15 seconds spanning the event time. Second, audio data is transcribed to text using the Automatic Speech Recognition model. Transcribed texts are then separated into utterance units using the utterance splitter. The DA classification model is applied to identify the intent of each utterance. Transcribed utterances within the MoI interval are then ordered by start time, and each adjacency pairs are modeled as an edge to construct a communication network. Each node in the communication network represents players.}
\end{figure*}
\section{Framework}
Figure \ref{fig:framework} presents an overview of the framework that automatically analyzes verbal communication patterns comprising three components: defining MoI, speech act recognition, and modeling communication networks.

\subsection{Defining Moments of Interest (MoI)}
Moments of Interest (MoI) is a conceptual framework to efficiently identify salient segments in continuous data streams~\cite{noroozi2024efficient}. 
We defined highly collaborative tasks as key game events (i.e., MoI) such as \texttt{Champion Kill}, \texttt{Elite Monster Kill} (e.g., Baron, Dragon), and \texttt{Building Destruction} (e.g., turret, inhibitor).
Previous study on communication analysis on MOBA games used a 30-second window~\cite{kim2024communication}, due to the fast-paced nature of the game. 
In accordance with prior studies, the MoI interval in this study was set to 30 seconds, spanning 15 seconds before and after the occurrence of each MoI. MoI intervals without verbal exchanges between players were excluded from the analysis.

A \texttt{Champion Kill} is defined as an event in which one player is eliminated with the involvement of at least three players, indicating a cooperative situation~\cite{mora2019team}.
\texttt{Champion Kills} occur throughout the game and provide a competitive advantage; thus, players actively cooperate to secure them.
Although solo kills can occasionally occur in the early stage of the game in one-on-one situations, the majority of kills involve the intervention of additional players (e.g., junglers or laners).
In such cases, players cooperate with teammates to incapacitate the opposing champion.
As the game progresses into the later stages, \texttt{Champion Kills} more frequently emerge from large-scale team fights involving multiple players.

Elite Monsters are neutral objectives that can be slain by either team to obtain valuable rewards, with each monster spawning at different times during the game.
For example, Baron Nashor spawns at 25 minutes and the team that secures it gains substantial advantages such as gold and powerful buffs.
However, Elite Monsters pose a significant challenge as they cannot be easily defeated by a single player and inflict considerable damage on champions.
Consequently, multiple players must coordinate to both prevent the opposing team from securing the objective and attempt to claim it for their own team, often resulting in large-scale team fights.

Buildings in LoL serve as key defensive structures and exert a critical influence on the outcome of the game. 
Accordingly, teams must defend their own buildings from destruction and seek to capitalize on advantageous situations to destroy the opponent’s. 
However, focusing solely on destroying the opponent's buildings exposes a team to counterattacks from the opponent. 
In such situations, teams must rapidly coordinate to decide whether to continue pressing the attack on the Buildings or to shift toward engaging the opposing team in a fight.

In these events, collaboration relies on communication through both information sharing and decision making. Within MoIs, team members exchange information to ensure a shared understanding of the game state. Players pose questions to seek relevant information (e.g., health status, spell availability) in order to assess the team’s current advantage or disadvantage. Other teammates respond by providing the requested information or by voluntarily contributing new information they have perceived.
In addition, MoIs frequently require team-level decisions, such as attempting to secure an Elite Monster or initiating a team fight. During this process, team members must negotiate their opinions, which typically unfolds when one player proposes a decision and the others respond by either accepting or rejecting the proposal.

\begin{table*}[t]
\centering
\caption{Transcription examples: The following dataset was obtained from an actual match of professional team A through our speech act recognition pipeline. The Dialogue Act (DA) labels are defined as follows: 0 = Inform, 1 = Question, 2 = Directive, 3 = Commissive.}
\label{tab:example-data}
\Description{Table 1: Table 1 describes dataset examples obtained from an actual match of professional team A through the speech act recognition model.}
\begin{tabularx}{\textwidth}{lXrrr}
\toprule
\textbf{Speaker} & \textbf{Text} & \textbf{Start (ms)} & \textbf{End (ms)} & \textbf{DA} \\
\midrule
Bot & \textit{If I take this lane completely, I'll probably hit six.} & 485106 & 486402 & 0 \\
Jungle & \textit{Can I push mid?} & 489032 & 491932 & 2 \\
Top & \textit{Yeah, just check the vision here for me while I’m heading over.} & 492338 & 510590 & 3 \\
Jungle & \textit{Yeah, I'm farming for Dragon, but they have ultimates, and we’ve already lost our Shen.} & 513076 & 523190 & 0 \\
Bot & \textit{I can also hit six.} & 526372 & 529236 & 0 \\
\bottomrule
\end{tabularx}
\end{table*}

\subsection{Speech Act Recognition}
\subsubsection{\textbf{Speech Transcription and Utterance Segmentation}}
We implemented a speech transcription pipeline to produce transcription with accurate timestamps in milliseconds.
The voice audio of all five players was collected separately, and the following procedures were applied individually to each recording.
We first transcribed the audio data to sentence-level text using Whisper large-v3~\cite{radford2023robust}, a state-of-the-art Automatic Speech Recognition (ASR) model that demonstrates high performance.
Timestamps indicating the start and end of each utterance were provided with a 10 ms resolution.
However, it is required to determine the start and end times of text with higher precision (i.e., 1 ms resolution) because in-game audio communication involves short utterances and fast-paced speech. 
Accordingly, we utilized a Connectionist Temporal Classification (CTC) model to achieve fine-grained alignment between audio and text with 1 ms resolution~\cite{graves2006connectionist}.
The CTC model takes long sequences of text and audio as input and outputs word-level text along with start and end timestamps for each word at 1 ms resolution. 
Then, the word-level segments were sequentially aggregated to reconstruct the utterance-level transcripts. 
Additionally, we employed a rule-based sentence-segmentation tool to divide the aggregated words into utterance-level transcripts. 
For each utterance, its temporal span was determined by aligning the onset of the first word and the offset of the final word. 
Subsequently, individual audio files from five players were transcribed following the same process. Each resulting dialogue was then translated into English using the Gemini 3:4b~\cite{team2023gemini} model to align with the requirements of the downstream dialogue act classification task.

\subsubsection{\textbf{Dialogue Act Classification}}
Dialogue Acts (DA) are higher-level semantic abstractions of utterances that convey the speaker's intent~\cite{searle2014speech}.
As proposed in He et al.~\cite{he2021speaker}, we combined utterance embeddings encoded from a pretrained language model RoBERTa~\cite{liu2019roberta} with speaker turn embeddings. These embeddings are then fed to the Bi-GRU~\cite{cho2014properties} model to infer DA. 
The model was trained using the DailyDialog (DYDA) dataset~\cite{li2017dailydialog} for 10 epochs with a learning rate of 0.0001 and dropout of 0.5. DYDA is a manually annotated corpus of dyadic, spontaneous conversations, and its four annotation labels (i.e., Inform, Question, Directive, and Commissive) effectively represent utterances at the illocutionary level~\cite{searle2014speech}. The Inform act encompasses utterances in which the speaker conveys information, including both assessments and opinions (e.g., Alice is at the bottom lane right now). The Questions act includes utterances where the speaker seeks to obtain information (e.g., Does Taliyah have flash?). The Directives Act comprises utterances such as requests, instructions, suggestions, and accept/reject offer (e.g., Let’s secure Baron). The Commissive category captures acts related to the acceptance or rejection of requests, suggestions, or offers (e.g., Directive: Let’s secure Baron. Commissive: Okay). Conceptually, the Inform and Questions categories pertain to information exchange, whereas the Directives and Commissive categories are associated with action-oriented negotiation~\cite{li2017dailydialog}.
As shown in Table~\ref{tab:example-data}, which presents data obtained from actual matches through our speech act recognition pipeline, the system demonstrates robust transcription performance even for domain-specific vocabulary related to the game.

\subsection{Modeling Communication Network}
\subsubsection{\textbf{Network Structure}}
We adopted SNA to quantitatively analyze utterances exchanged between players during conversations~\cite{tabassum2018social}, as networks are considered a valid tool to investigate complex dynamics in team sports~\cite{passos2011networks}. 
The communication network was constructed with five nodes, each corresponding to one of the players participating in a LoL match. 
\begin{figure*}
\centering
  \includegraphics[width=0.5\linewidth]{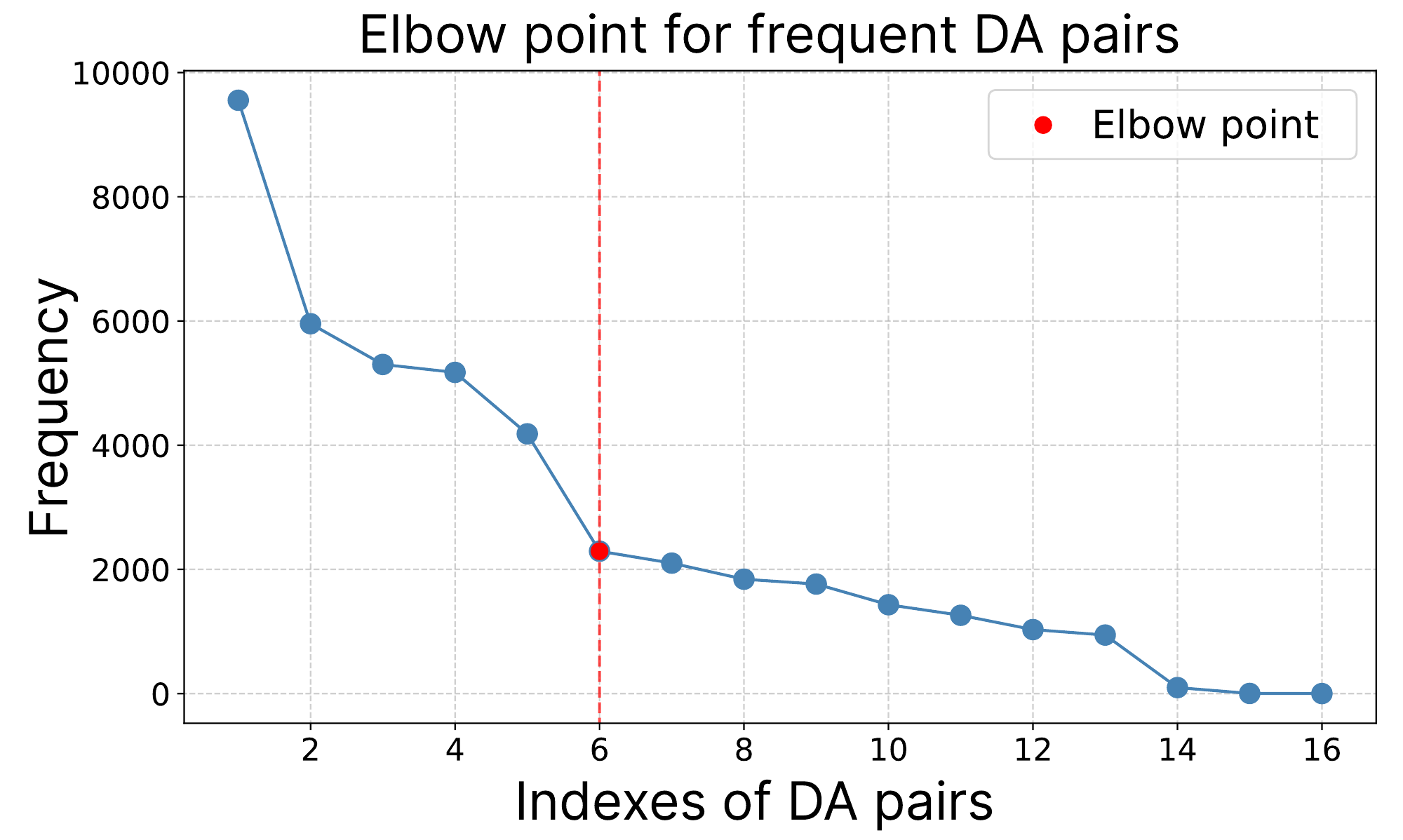}
  \caption{Line graph indicating the elbow point for frequent Dialogue Act (DA) pairs. The x-axis represents the indices of DA pairs sorted in descending order of frequency, and the y-axis shows the frequency of each DA pair. Using the Kneedle algorithm, we identified six DA pairs where additional increases became negligible.}
  \label{fig:kneedle}
  \Description{Figure 3: Figure 3 is a line graph that illustrates the elbow point for determining the number of frequent DA pairs to include in our analysis. The additional increases become negligible from index 6, which is the reason why we included six DA pairs in our study.}
\end{figure*}
Within each MoI time window (i.e., 30 seconds), utterance sequences were represented as consecutive adjacency pairs.
For each pair, the first utterance was defined as the sender’s utterance, and the second utterance as the receiver’s utterance. 
The sender and receiver were determined in the order of utterance start times, with each utterance defined as the sender’s utterance and the immediately following utterance defined as the receiver’s utterance~\cite{schegloff1973opening}.
This procedure was applied sequentially across the entire utterance sequence, such that every consecutive pair of utterances was modeled as a directed edge from the sender to the receiver. If the time interval between two utterances (i.e., pairing interval) exceeded five seconds, the pair was not considered a conversational exchange between the two players and was therefore excluded from edge construction.
When the same type of utterance between two players is repeated, the corresponding edge weight increases.
Note that the set of nodes remains fixed as no player substitutions occur during a match, and certain nodes may exhibit no incident edges during a given interval if the corresponding player remained verbally inactive.

\begin{figure*}
\centering
  \includegraphics[width=1\linewidth]{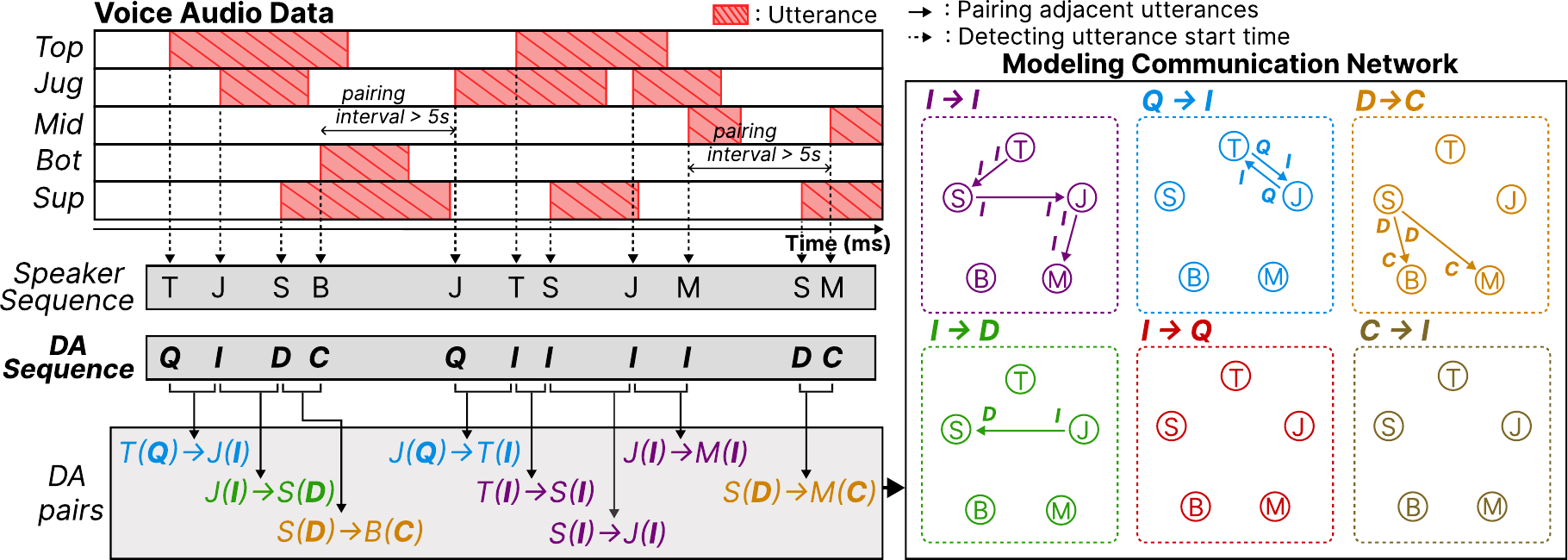}
  \caption{An example of modeling a communication network. Utterance sequences are divided into pairs, which are modeled as directed edges from the sender to the receiver, and nodes are modeled as players. If the time interval between the start time of adjacent utterances (i.e., pairing interval) is larger than 5 seconds, such adjacent pairs are excluded from network formation. T, J, M, B, S denote Top, Jungle, Mid, Bot, Support, respectively. Dialogue Act (DA) I, Q, D, C each represent Inform, Question, Directive, Commissive. }
\label{fig:network}
\Description{Figure 4: Figure 4 depicts an example of modeling a communication network. In this figure, the speaker and the corresponding DA are ordered by start time, and are described as the player role (DA). After obtaining speaker and DA sequences, adjacent DA pairs are constructed as an edge if the interval between the two utterances' start times is less than 5 seconds. Communication networks are constructed separately using utterances corresponding to six frequent DA pairs: Inform → Inform, Question → Inform, Directive → Inform, Inform → Directive, Inform → Question, Commissive → Inform.}
\end{figure*}

In conversations, there are constraints on how one utterance should be followed by another, forming recognizable patterns that create normative obligation for a response~\cite{sacks1974simplest}.
Similarly, utterances may follow recurrent sequential patterns rather than occurring randomly. 
To identify such frequently emerging DA patterns, we applied T-pattern detection~\cite{magnusson2000discovering}.
 Additionally, we employed the Kneedle algorithm~\cite{satopaa2011finding}, which is commonly used to detect the point at which the increase in occurrence counts begins to level off to determine the optimal number of DA pairs to include in our analysis. As shown in Fig.~\ref{fig:kneedle}, the line graph of DA pair frequencies indicates that the rate of change markedly decreases after the sixth pair. This point was considered the elbow, suggesting that additional increases are no longer meaningful. Accordingly, we focused on the six most frequent patterns: Inform $\rightarrow$ Inform, Question $\rightarrow$ Inform, Directive $\rightarrow$ Commissive, Inform $\rightarrow$ Directive, Inform $\rightarrow$ Question, and Commissive $\rightarrow$ Inform. Networks are constructed separately for utterances corresponding to each DA pair, and the resulting networks were analyzed through network metrics to investigate how communication patterns differed across pairs.
An example of the resulting network is represented in Fig. \ref{fig:network}.
We implemented the network using the Python package NetworkX~\cite{SciPyProceedings_11}.

\subsubsection{\textbf{Metrics}}
We calculated three standard SNA measures to characterize the structural patterns of communication flow: network density, indegree centralization, and outdegree centralization~\cite{sparrowe2001social}. 
Network density is defined as the proportion of actual connections in the network relative to maximum possible connections, which explains the level of connectedness in a network~\cite{borgatti2003relational}. 

The density is calculated based only on the presence or absence of communication between nodes. 
In terms of communication analysis, if any conversation occurred between two nodes, an edge is considered to exist between them, regardless of its frequency. 
Density is calculated as:
\[
\rho = \frac{m(G)}{m_{\max}(G)}
\]
where $m(G)$ represents the actual number of edges in the network, and $m_{\max}(G)$ indicates the maximum possible number of edges. $m_{\max}(G)$ equals N(N-1)/2, where N denotes the number of nodes (i.e., 5 players in our case).

Network centralization assesses the extent of a network’s dispersion or inequality among all actors (e.g., speakers in communication) prominences~\cite{knoke2008social}.
In the context of communication networks, an individual actor’s prominence reflects its greater visibility to the other network actors, which indicates high involvement in many relations~\cite{knoke2008social}. In a directed weighted communication network, an actor’s prominence can be measured by outdegree centrality $c_{OD}(i)$, defined as the sum of the weights of outgoing edges (i.e., senders' utterance), and indegree centrality $c_{ID}(i)$, defined as the sum of the weights of incoming arcs (i.e., receivers' uttterance).
Following Freeman’s study based on differences in point centralities~\cite{freeman1978centrality}, network centralization is defined as follows:
\begin{gather}
c_{OD}(i)=\sum_{j=1}^{N} w_{ij}, \qquad
c_{ID}(i)=\sum_{j=1}^{N} w_{ji}, \tag{1} \\[6pt]
C_{OD} = \frac{\sum_{i=1}^{N}(c_{OD}^{\max} - c_{OD}(i))}{(N-1)U}, \qquad
C_{ID} = \frac{\sum_{i=1}^{N}(c_{ID}^{\max} - c_{ID}(i))}{(N-1)U} \tag{2}
\end{gather}

$w_{ij}$ refers to edge weights from player \textit{i} to \textit{j} during a certain interval. Both $c^{\max}$ are the largest observed indegree and outdegree centralities, respectively. U corresponds to the number of utterances during the interval, and the denominator (i.e., $(N-1)U$) reflects the centralization when all utterances are concentrated through a single individual, thus standardizing the result. 
\begin{figure*}
\centering
  \includegraphics[width=1\linewidth]{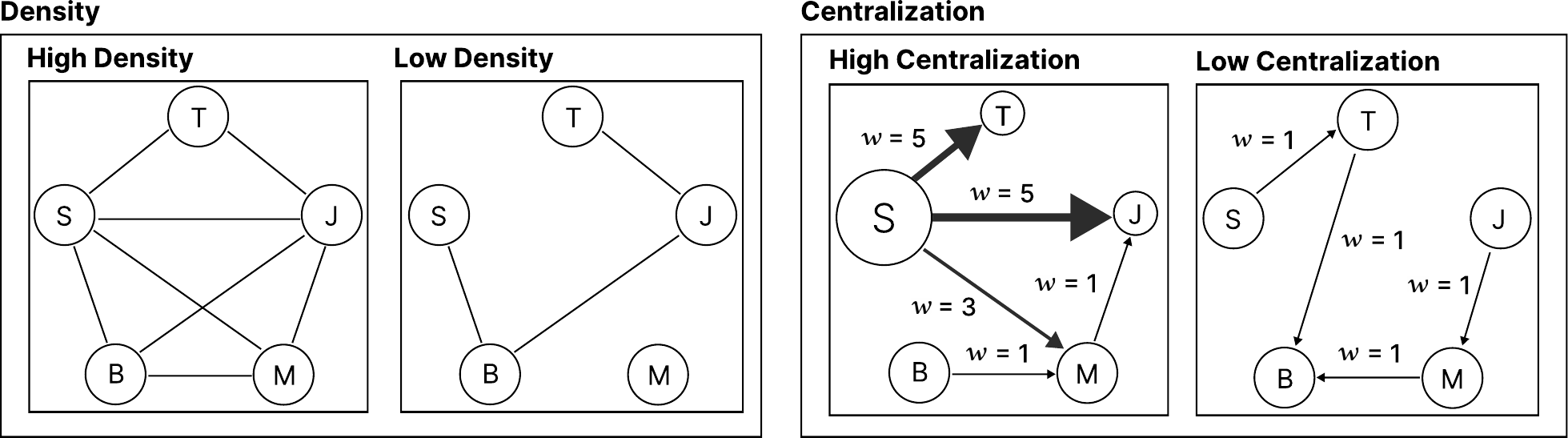}
  \caption{Illustration of communication networks contrasting properties: Networks with high density indicate broad participation among players, while low-density networks show sparse interactions; networks with high centralization are dominated by a few nodes, whereas networks with low centralization exhibit more balanced participation. Node and edge size in the right panel (centralization) are proportional to its centrality and weights (i.e., $w$), respectively.}
  \label{fig:metric}
  \Description{Figure 5: Figure 5 illustrates examples of 4 networks of varying density and centralization. All players except Top and Bot, Top and Mid are connected in the high-density network, whereas only Top and Jungle, Jungle and Bot, Bot and Jungle are connected in the low-density network. Two networks with high and low centralization are depicted in the right panel. In this panel, node and edge size correspond to centrality and weights, respectively. In the highly centralized network, the player with the Support role dominated the conversation, having the highest outgoing edge weights, whereas all players communicated once in the network with low centralization.}
\end{figure*}
Examples of communication networks with varying density and centralization values are shown in Fig. \ref{fig:metric}. In the left figure, two networks differing in density are shown. 
If every player had spoken to at least one other player, resulting in active communication, the network would have resembled the high-density case. 
In the example of high density in Fig. \ref{fig:metric}, 8 of the 10 possible edges are present, resulting density equals 0.8. 
By contrast, in the low-density example, utterances occurred only between three pairs of players, producing a density of 0.3.
The right panel depicts networks with different levels of outdegree centralization. Node size corresponds to each player’s centrality and edge thickness is proportional to its weight. The player with the support role has the largest number and weight of outgoing edges in the left network, indicating that this player dominated the conversation, thus resulting in high outdegree centralization. In contrast, each player produced one outgoing edge of equal weight in the network on the right, which indicates low outdegree centralization. Indegree centralization can be computed analogously.

\section{Data Acquisition}
\subsection{Participants}
We collected data from 25 participants (15 professional LoL players from three teams and 10 amateur LoL players from two teams), who had sufficient experience in LoL during the current year. 
We recruited teams consisting of professional players who had received professional training on team communication and had competed in the national championship at least five times. Amateur players recreationally played LoL flex queues and had never received training regarding communication.
Each participant belonged to one of the teams that had been formed before data collection and regularly played the game together. In order to mitigate the differences in team familiarity and trust that may arise from varying durations of team formation~\cite{ching2024competitive}, we recruited only teams that had been formed for less than one month.
To minimize the impact of the game experience or skill on communication, we tried to recruit participants with similar levels of expertise. 
Therefore, we only recruited players who had played LoL for at least 4 years and whose average rank is Master (top 0.5\%) or higher. 
The average tier of professional teams is grandmaster (top 0.051\%), and master for amateur teams. 
Also, to qualitatively validate the analysis results based on domain knowledge, we recruited four experts from a professional esports team, including two coaches, one manager, and an esports-data analyst.

The average amount of LoL playing time per week was 49.4 hours (SD = 29.77) for the professional teams and 15.4 hours (SD = 6.64) for amateur teams. The average age of the professional players was 18.8 years old (SD = 2.01) and 22.2 years old (SD = 3.29) for the amateur players. 
The professional players had an average of 7.0 years of LoL experience (SD = 2.75), whereas the amateur players reported a slightly longer average of 8.3 years (SD = 2.98).

\subsection{Setup and Procedure}
Prior to data collection, all players were required to enter the Discord server where Craig and the Checker Bot were installed. We instructed players to adjust Discord voice sensitivity settings to prevent background or noise from interfering with voice data. 
When the game started, the Craig Bot detected and recorded the audio of all players in the channel, and the Checker Bot recorded the start time of the recording. 
During gameplay, players communicated freely while wearing headsets. 
At the end of each game, the recorded audio data from all players was automatically uploaded to Google Drive. 
The recording began automatically with the start of the game until at least one player left the channel with the Craig Bot.

\subsection{Dataset Description}
A total of 89 games were collected. Since the earliest possible time a match can conclude is 15 minutes, 4 games that ended before this point were excluded from the analysis.
Those games are considered either forced termination or system errors. 
As a result, 84 games were included in the dataset, comprising 41 games from professional team A, 11 games from professional team B, 13 games from professional team C, 10 games from amateur team A, and 9 games from amateur team B. Professional team A contributed the largest number of games because they conducted scrimmages more frequently each week.
The mean game duration across all matches was 27.5 minutes (SD = 5.33). When broken down by team, professional team A exhibited the longest average game duration (M = 29.50, SD = 5.87), followed by professional team B (M = 28.88, SD = 5.73), professional team C (M = 27.5, SD = 5.33), amateur team B (M = 28.35, SD = 7.33), and amateur team A with the shortest duration (M = 26.42, SD = 5.6).

Across all games, a total of 45,660 utterances were recorded, with an average of 537.2 utterances per game (SD = 188.7). Among the teams, professional team A showed the highest average number of utterances (M = 648.8, SD = 157.15), followed by amateur team B (M = 505.8, SD = 205.13), professional team C (M = 495.1, SD = 101.89), professional team B (M = 384.7, SD = 83.1), and amateur team A (M = 333.2, SD = 129.13).

\subsection{Speech Act Recognition Pipeline Evaluation}
\yc{To evaluate the performance of our framework’s speech act recognition pipeline, we manually labeled the transcriptions and DAs of 10 games (two per team) from our dataset. The transcripts were annotated by the second author, who has three years of professional esports coaching experience. After familiarizing with the examples of each DA category in the DYDA dataset, the annotator assigned each utterance to one of the four DA types. In total, we transcribed 6,260 utterances, consisting of 3,130 Inform, 672 Question, 1,651 Directive, and 772 Commissive acts.}

\yc{Word Error Rate (WER) is most commonly employed to evaluate ASR models, defined as the sum of the number of substituted words, the number of deletions, and the number of insertions divided by the total number of words in the ground truth transcription. Whisper large-v3 yielded a WER of 1.87 on our dataset without fine-tuning. While this value is higher than the reported WERs of below 0.6 on public benchmarks such as Common Voice 15~\cite{ardila2019common} and FLEURS~\cite{conneau2023fleurs}, it is acceptable given the high-pressure, rapid exchanges characteristic of our dataset. Finally, when training our DA classification model under the conditions described in Section 4.2.2, the model achieved an accuracy of 57.6\% on our dataset.}
\begin{table*}
\centering
\resizebox{\textwidth}{!}{%
\setlength{\tabcolsep}{4pt}
\begin{tabular}{lcccccccccccc}
\toprule
& \multicolumn{4}{c}{$\rho$} & \multicolumn{4}{c}{$C_{OD}$} & \multicolumn{4}{c}{$C_{ID}$} \\
& \multicolumn{1}{c}{\begin{tabular}[c]{@{}c@{}}MoI\\[-1pt]M (SD)\end{tabular}} 
& \multicolumn{1}{c}{\begin{tabular}[c]{@{}c@{}}Non-MoI\\[-1pt]M (SD)\end{tabular}} 
& $W$ & $p$ 
& \multicolumn{1}{c}{\begin{tabular}[c]{@{}c@{}}MoI\\[-1pt]M (SD)\end{tabular}} 
& \multicolumn{1}{c}{\begin{tabular}[c]{@{}c@{}}Non-MoI\\[-1pt]M (SD)\end{tabular}} 
& $W$ & $p$ 
& \multicolumn{1}{c}{\begin{tabular}[c]{@{}c@{}}MoI\\[-1pt]M (SD)\end{tabular}} 
& \multicolumn{1}{c}{\begin{tabular}[c]{@{}c@{}}Non-MoI\\[-1pt]M (SD)\end{tabular}} 
& $W$ & $p$ \\
\cmidrule(lr){2-5} \cmidrule(lr){6-9} \cmidrule(lr){10-13}
I $\rightarrow$ I & 0.166 (0.101) & 0.163 (0.092) & 116319 & 0.719 
    & 0.169 (0.096) & 0.162 (0.103) & 155671 & 0.068 
    & 0.166 (0.097) & 0.160 (0.106) & 161896 & 0.087 \\
Q $\rightarrow$ I & \textbf{0.145 (0.076)} & 0.131 (0.063) & 54169 & *** 
    & 0.175 (0.099) & \textbf{0.188 (0.095)} & 98114 & *** 
    & 0.176 (0.099) & \textbf{0.188 (0.095)} & 102020 & *** \\
D $\rightarrow$ C & \textbf{0.161 (0.099)} & 0.140 (0.072) & 32899 & *** 
    & 0.177 (0.098) & \textbf{0.187 (0.098)} & 52320 & *** 
    & 0.173 (0.101) & \textbf{0.193 (0.094)} & 40866 & 0.067 \\
I $\rightarrow$ D & 0.148 (0.081) & 0.140 (0.093) & 27241 & 0.061
    & 0.181 (0.093) & 0.181 (0.098) & 54967 & 0.740 
    & 0.180 (0.094) & 0.173 (0.101) & 61923 & 0.120 \\
I $\rightarrow$ Q & 0.135 (0.065) & 0.129 (0.060) & 10660 & 0.080 
    & 0.192 (0.093) & 0.195 (0.090) & 24361 & 0.895 
    & \textbf{0.195 (0.090)} & 0.186 (0.096) & 21048 & 0.021* \\
C $\rightarrow$ I & \textbf{0.122 (0.054)} & 0.112 (0.033) & 754 & 0.027* 
    & 0.213 (0.077) & 0.227 (0.067) & 1261 & 0.079 
    & 0.213 (0.077) & 0.209 (0.088) & 2005 & 0.296 \\
\bottomrule
\Description{Table 2: Results on comparison of network metrics between Moments of Interest (MoI) and non-MoI are shown in Table 2. Communication networks demonstrated significantly higher density when using utterances corresponding to Question → Inform, Directive → Commissive, Inform → Question pairs. Moreover, centralization was lower during MoI when networks were constructed using utterances of Question → Inform, Directive → Commissive, with the only exception being indegree centralization in the Directive → Commissive pair. In contrast, indegree centralization was higher during MoI in Inform → Question pairs.}
\end{tabular}}
\caption{Comparison of network indicators (Density $\rho$, Outdegree Centralization $C_{OD}$, and Indegree Centralization $C_{ID}$) between MoI and non-MoI across utterances corresponding to frequent Dialogue Act (DA) pairs. I, Q, D, and C denote Inform, Question, Directive, and Commissive, respectively. Higher values are highlighted for metrics that showed significant differences between MoI and non-MoI. Values are presented as Mean (SD); Wilcoxon signed-rank test statistics $W$ and $p$-values are reported (* $p<.05$, ** $p<.01$, *** $p<.001$).}
\label{tab:moiresults}
\end{table*}
\section{Data Analysis} %

\subsection{\textbf{Preprocessing}}
\subsubsection{\textbf{Paring MoI and non-MoI}}
In order to investigate differences in team communication patterns under conditions requiring cooperation, we first defined situations requiring cooperation as MoIs and situations not requiring cooperation as non-MoIs.
It is essential to appropriately pair each MoI with a corresponding non-MoI to enable statistically valid comparisons between these two conditions.
For example, pairing an MoI with a non-MoI that is separated by a large temporal gap may yield an inappropriate basis for comparison.
Because team dynamics, game state, and communication intensity may change considerably over time, such variation can confound the observed differences in communication~\cite{lee2025less}. 

To minimize the potential influences of the situational factors, we paired each MoI with the nearest non-MoI.
First, the corresponding non-MoI was initially defined as the 30-second interval preceding the MoI onset (i.e., from 30 seconds before the MoI start time); if this window overlapped with another MoI or did not allow network construction (e.g., no utterances or only a single speaker), the non-MoI window was shifted earlier by one second until a valid interval was identified.

\subsubsection{\textbf{Segmentation of Game Phases based on Changes in Cooperation}}
Based on the interview with four esports experts, we divided the game into four distinct phases reflecting different cooperative demands. Expert consultation identified key transition points at 5 minutes (first Elite Monster spawn), 14 minutes (turret plate removal), and 25 minutes (Baron Nashor availability) where the nature of required team coordination fundamentally shifts.

The early laning phase (0-5 minutes) emphasizes information-centric cooperation, with players sharing intelligence about opponent positioning and potential threats while maintaining individual lane focus. The late laning phase (5-14 minutes) introduces objective-oriented cooperation as contestable Elite Monsters emerge, requiring teams to balance individual laning with collective decision-making about map-wide objectives. The team fight phase (14-25 minutes) represents peak collaborative complexity, as turret plate removal enables fluid cross-map movement and frequent multi-player engagements that often determine game outcomes. The endgame phase (25+ minutes) features high-stakes coordination around Baron Nashor, typically characterized by asymmetric team states where leading teams apply coordinated pressure while trailing teams attempt synchronized comeback strategies.

\subsection{Results}
\subsubsection{\textbf{Descriptive Statistics}}
Across all games, a total of 5,517 MoIs were identified, corresponding to an average of 46.00 per game (SD = 17.96). 
Professional team A exhibited the highest average number of MoIs (M = 63.51, SD = 17.71), followed by professional team C (M = 49.54, SD = 14.01), amateur team B (M = 53.00, SD = 20.33), amateur team A (M = 41.40, SD = 22.38), and professional team B with the lowest mean (M = 38.45, SD = 17.96).
With respect to key game events, 3,777 MoIs were associated with \texttt{Champion Kills} (M = 44.96 per game, SD = 14.23), 766 with \texttt{Elite Monster Kills} (M = 9.12 per game, SD = 1.75), and 974 with \texttt{Building Destruction} (M = 11.60 per game, SD = 4.63).
The temporal distribution of MoIs revealed a tendency for their frequency to increase as games progressed: 230 occurred during the early laning phase, 1,437 during the late laning phase, 2,590 during teamfights, and 1,260 during the endgame.

Application of the DA classifier yielded 20,291 Inform, 9,626 Question, 10,060 Directive, and 5,481 Commissive utterances. Across all MoIs, the number of MoIs in which DA pairs occurred between two different players was 1,232 for Inform → Inform, 1,130 for Question → Inform, 820 for Directive → Commissive, 819 for Inform → Directive, 626 for Inform → Question, and 269 for Commissive → Inform.

\subsubsection{\textbf{H1: Verbal Communication Pattern in Situations Requiring Collaboration}}
We compared verbal communication patterns between situations requiring collaboration (i.e., MoIs) and those not requiring collaboration (i.e., non-MoIs). 
For each MoI and non-MoI interval, a communication network was constructed based on the utterances that occurred within the interval, and the communication patterns were quantified using three metrics: density, indegree centralization, and outdegree centralization.
We applied the Wilcoxon signed-rank test for within-subject comparisons of metrics between MoI and non-MoI, as our samples are not normally distributed.

As shown in Table \ref{tab:moiresults}, differences in metrics between MoI and non-MoI situations emerged only for certain DA pairs when communication networks were constructed.. 
Networks built from Question → Inform and Directive → Commissive pairs exhibited significantly higher density during MoI (p < .001), indicating broader participation of players in these types of exchanges during moments requiring collaboration. 
Commissive → Inform pairs displayed a similar tendency, although the effect was weaker (p < .05).
Centralization showed significantly lower values during MoI for both Question → Inform and Directive → Commissive pairs, with the only exception being indegree centralization in the Directive → Commissive pair (p < .001 for all other cases). 
Inform → Question pairs exhibited the opposite trend, with significantly lower centralization during non-MoI (p < .05). 
Inform → Inform pairs occurred most frequently due to the large proportion of Inform acts in the overall dataset, but this pair revealed no significant differences between MoI and non-MoI in either density or centralization. Given the significant differences between collaborative situations, H1 is supported.

\begin{figure*}
    \centering
    \includegraphics[width=1\linewidth]{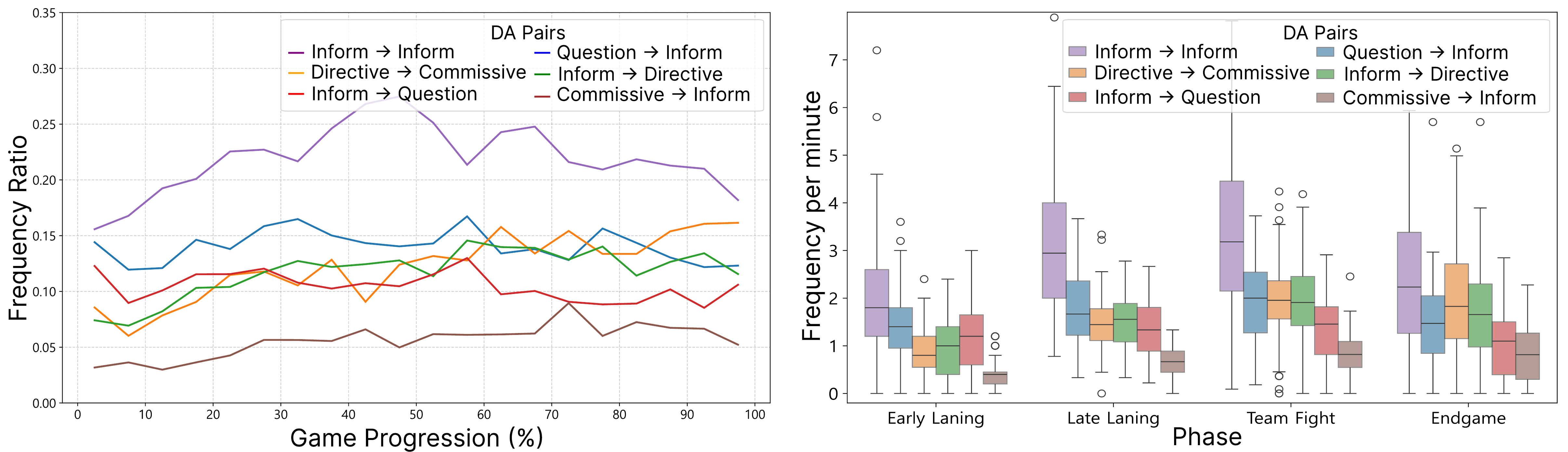}
    \caption{Temporal trends in the relative frequency of Dialogue Act (DA) pairs across game progression (left), and distribution of DA pair frequencies per minute across game phases (right). The line graph presents the game progress by normalizing elapsed time with the total game duration. For each 1\% interval of game progress, we calculated the ratio of each DA pair as the frequency of occurrences within 30 seconds of that interval divided by the total number of pairs. To smooth local fluctuations and highlight broader trends, a 5\% binning procedure was applied. The box plot shows the number of DA pairs occurring within each game phase. These counts were normalized by the duration of each phase (in minutes) across the dataset, yielding the frequency of DA pairs per minute.}
    \label{fig:two_boxplots}
    \Description{Figure 6: Figure 6 depicts temporal trends of six frequent DA pairs using both line graphs and box plots. Inform → Inform pair is the most frequent DA pair during the game, reflecting its large proportion of Inform acts in the overall datasets. Directive → Commissive pair shows the sharpest rise, starting as the second least frequent pair but becoming the second most common in the endgame phase. Question → Inform and Inform → Directive pairs show a moderate increase in the beginning, but ultimately decrease in the endgame. Other pairs display relatively stable patterns.}
\end{figure*}

\subsubsection{\textbf{H2: Verbal Communication Pattern across Game Phase}}
LoL is a game in which the type of cooperation required changes over time.
According to expert interviews we conducted, team cooperation in the game changes around the 14-minute mark. 
Before 14 minutes, gameplay is primarily classified as the laning phase, where cooperation largely involves sharing knowledge about the information related to in-game contexts. 
When each lane's turret plates that grant additional gold upon destruction disappear at 14 minutes, the importance of laning diminishes; players are no longer confined to their individual lanes, and team-level coordination becomes increasingly critical. Therefore, we investigated how changes in cooperation correspond to changes in communication patterns over time by visualizing the frequency of each DA pair across game progression and phase.

Figure~\ref{fig:two_boxplots} presents the distribution of DA pairs across game progression, illustrated with both line graphs and box plots. The total frequency of utterances increased as the game advanced, yet the trajectories of individual DA pairs diverged. 
The Directive → Commissive pair exhibited a consistent increase as the game progressed. It was among the least frequent DA pairs at the beginning but ultimately ranked as the second most common pair in the endgame phase, surpassed only by the Inform → Inform pair.
The prevalence of Inform → Inform pairs reflects the overall dominance of Inform acts; however, this pair exhibited a different trajectory, increasing until the team fight phase and then declining toward the endgame. 
Question → Inform and Inform → Directive pairs also showed moderate growth in the early stages but gradually decreased as the game progressed. 
In contrast, Inform → Question and Commissive → Inform pairs displayed relatively stable patterns, with little variation across phases.
Because Directive → Commissive has shown a significant increase as the game progressed, H2 holds true.

\subsubsection{\textbf{H3: Verbal Communication Pattern across Different Teams}}
All three professional teams were professionally trained, and their coaches reported providing feedback on the teams’ communication practices. 
For example, players were instructed by their coaches to share perceived information immediately and to actively issue directives without being constrained by their specific roles.
We investigated whether trained teams exhibited different communication patterns during situations requiring collaboration. 
For this purpose, we compared communication network metrics from utterances produced during MoIs for each group using the Mann-Whitney U test.

\begin{table*}
\centering
\resizebox{\textwidth}{!}{%
\small
\setlength{\tabcolsep}{4pt}
\begin{tabular}{lcccccccccccc}
\toprule
& \multicolumn{4}{c}{$\rho$} 
& \multicolumn{4}{c}{$C_{OD}$} 
& \multicolumn{4}{c}{$C_{ID}$} \\
& \multicolumn{1}{c}{\begin{tabular}[c]{@{}c@{}}Pro\\[-1pt]M (SD)\end{tabular}} 
& \multicolumn{1}{c}{\begin{tabular}[c]{@{}c@{}}Amateur\\[-1pt]M (SD)\end{tabular}} 
& $U$ & $p$ 
& \multicolumn{1}{c}{\begin{tabular}[c]{@{}c@{}}Pro\\[-1pt]M (SD)\end{tabular}} 
& \multicolumn{1}{c}{\begin{tabular}[c]{@{}c@{}}Amateur\\[-1pt]M (SD)\end{tabular}} 
& $U$ & $p$ 
& \multicolumn{1}{c}{\begin{tabular}[c]{@{}c@{}}Pro\\[-1pt]M (SD)\end{tabular}} 
& \multicolumn{1}{c}{\begin{tabular}[c]{@{}c@{}}Amateur\\[-1pt]M (SD)\end{tabular}} 
& $U$ & $p$ \\
\cmidrule(lr){2-5} \cmidrule(lr){6-9} \cmidrule(lr){10-13}
I $\rightarrow$ I
& 0.138 (0.073) & 0.123 (0.044) & 118593 & 0.052
& 0.176 (0.101) & 0.192 (0.089) & 102771 & 0.054
& 0.176 (0.099) & 0.190 (0.091) & 111610 & 0.101 \\
Q $\rightarrow$ I
& \textbf{0.129 (0.059)} & 0.119 (0.047) & 117766 & 0.028* 
& 0.198 (0.086) & 0.200 (0.091) & 107713 & 0.487
& 0.195 (0.089) & 0.200 (0.091) & 114654 & 0.290 \\
D $\rightarrow$ C
& \textbf{0.123 (0.054)} & 0.109 (0.031) & 47612 & 0.005** 
& 0.196 (0.094) & \textbf{0.214 (0.083)} & 39632 & 0.025*
& 0.191 (0.098) & \textbf{0.217 (0.079)} & 37936 & 0.002** \\
I $\rightarrow$ D
& \textbf{0.122 (0.049)} & 0.109 (0.032) & 68196 & 0.001** 
& 0.199 (0.089) & 0.207 (0.089) & 57824 & 0.127
& 0.200 (0.089) & 0.207 (0.089) & 57994 & 0.144 \\
I $\rightarrow$ Q
& 0.120 (0.048) & 0.111 (0.031) & 49583 & 0.056
& 0.205 (0.086) & 0.221 (0.069) & 42589 & 0.069
& 0.201 (0.088) & \textbf{0.220 (0.072)} & 42303 & 0.003** \\
C $\rightarrow$ I
& 0.113 (0.041) & 0.106 (0.025) & 14913 & 0.360
& 0.217 (0.077) & 0.219 (0.076) & 14230 & 0.761
& 0.213 (0.081) & 0.219 (0.076) & 13984 & 0.518 \\
\bottomrule
\end{tabular}}
\caption{Comparison of $\rho$, $C_{OD}$, $C_{ID}$ between professional and amateur teams using Mann-Whitney U test. I, Q, D, and C denote Inform, Question, Directive, Commissive, respectively. Higher values in density and centralization that are significantly different across professional and amateur teams are boldfaced. Values are presented as Mean (SD), with corresponding $U$ statistics and $p$-values (* $p<.05$, ** $p<.01$, *** $p<.001$).}
\label{tab:team_mann}
\Description{Table 3: Comparison result from the Mann-Whitney U test between professional and amateur teams. Professional teams exhibited higher density and lower centralization, but significant differences on both metrics were only observed in Directive → Commissive pairs.}
\end{table*}

The results of the between-group comparison using the Mann–Whitney U test are presented in Table \ref{tab:team_mann}. 
Professional teams exhibited higher network density than amateur teams, although significant differences were observed only for the Question → Inform, Directive → Commissive, and Inform → Directive pairs (p < .05). Centralization measures were generally lower in professional teams, yet significant effects emerged only for indegree and outdegree centralization of the Directive → Commissive pair  (all p < .05) and for indegree centralization of the Inform → Question pair (p < .01).

Although professional teams and amateur teams were matched in terms of experience and competitive tier, their communication practice can be affected by non-skill-related factors such as dependence on specific players or communication effectiveness.
Therefore, it is necessary to examine how communication patterns differed across teams and to investigate the underlying reasons for these differences.
Network metrics derived from utterances during MoI were compared among teams using the Kruskal-Wallis test, followed by the pairwise Mann-Whitney U test as post hoc to identify specific between-team differences.

\begin{table*}[t]
\centering
\begin{tabular}{c c c c c c c c c}
\toprule
\multirow{2}{*}{Metric} & \multirow{2}{*}{Pair} 
 & Pro A & Pro B & Pro C & Amateur A & Amateur B 
 & \multirow{2}{*}{$H$} & \multirow{2}{*}{$p$} \\
 &  & M (SD) & M (SD) & M (SD) & M (SD) & M (SD) &  &  \\
\midrule
\multirow{6}{*}{$\rho$} 
 & I $\rightarrow$ I & \textbf{0.175 (0.105)} & 0.127 (0.057) & 0.138 (0.063) & 0.133 (0.069) & 0.144 (0.068) & 86.587 & *** \\
 & Q $\rightarrow$ I & \textbf{0.160 (0.091)} & 0.106 (0.027) & 0.145 (0.073) & 0.129 (0.055) & 0.131 (0.056) & 81.488 & *** \\
 & D $\rightarrow$ C & \textbf{0.163 (0.101)} & 0.119 (0.048) & 0.143 (0.076) & 0.117 (0.049) & 0.138 (0.076) & 62.696 & *** \\
 & I $\rightarrow$ D & \textbf{0.151 (0.082)} & 0.114 (0.048) & 0.128 (0.059) & 0.111 (0.031) & 0.130 (0.065) & 86.308 & *** \\
 & I $\rightarrow$ Q & \textbf{0.143 (0.074)} & 0.121 (0.055) & 0.128 (0.061) & 0.130 (0.062) & 0.132 (0.061) & 20.312 & *** \\
 & C $\rightarrow$ I & \textbf{0.129 (0.064)} & 0.114 (0.048) & 0.111 (0.034) & 0.102 (0.014) & 0.114 (0.043) & 25.280 & *** \\
\cmidrule(lr){2-9}
 & Average & \textbf{0.156} & 0.117 & 0.134 & 0.124 & 0.133 &  &  \\
\midrule
\multirow{6}{*}{$C_{OD}$} 
 & I $\rightarrow$ I & 0.162 (0.098) & 0.175 (0.103) & 0.181 (0.099) & \textbf{0.191 (0.091)} & 0.174 (0.097) & 26.033 & *** \\
 & Q $\rightarrow$ I & 0.171 (0.099) & \textbf{0.213 (0.086)} & 0.193 (0.088) & 0.192 (0.094) & 0.190 (0.089) & 45.471 & *** \\
 & D $\rightarrow$ C & 0.174 (0.099) & 0.218 (0.073) & 0.189 (0.096) & \textbf{0.221 (0.076)} & 0.184 (0.099) & 54.968 & *** \\
 & I $\rightarrow$ D & 0.179 (0.095) & \textbf{0.222 (0.073)} & 0.200 (0.087) & 0.201 (0.094) & 0.192 (0.096) & 53.328 & *** \\
 & I $\rightarrow$ Q & 0.181 (0.097) & \textbf{0.211 (0.089)} & 0.196 (0.095) & 0.201 (0.089) & 0.186 (0.098) & 21.411 & *** \\
 & C $\rightarrow$ I & 0.205 (0.084) & 0.221 (0.075) & 0.226 (0.068) & \textbf{0.250 (0.054)} & 0.210 (0.084) & 27.026 & *** \\
\cmidrule(lr){2-9}
 & Average & 0.176 & \textbf{0.209} & 0.194 & 0.202 & 0.188 &  &  \\
\midrule
\multirow{6}{*}{$C_{ID}$} 
 & I $\rightarrow$ I & 0.161 (0.098) & 0.178 (0.101) & 0.184 (0.097) & \textbf{0.188 (0.091)} & 0.164 (0.101) & 27.855 & *** \\
 & Q $\rightarrow$ I & 0.168 (0.100) & \textbf{0.209 (0.089)} & 0.186 (0.093) & 0.186 (0.098) & 0.190 (0.089) & 40.273 & *** \\
 & D $\rightarrow$ C & 0.172 (0.100) & 0.218 (0.073) & 0.184 (0.098) & \textbf{0.219 (0.080)} & 0.189 (0.096) & 56.470 & *** \\
 & I $\rightarrow$ D & 0.177 (0.096) & \textbf{0.230 (0.062)} & 0.194 (0.092) & 0.195 (0.099) & 0.200 (0.090) & 67.012 & *** \\
 & I $\rightarrow$ Q & 0.182 (0.097) & \textbf{0.213 (0.086)} & 0.204 (0.086) & 0.210 (0.082) & 0.185 (0.098) & 31.403 & *** \\
 & C $\rightarrow$ I & 0.204 (0.085) & 0.224 (0.072) & 0.230 (0.062) & \textbf{0.245 (0.035)} & 0.206 (0.088) & 29.861 & *** \\
\cmidrule(lr){2-9}
 & Average & 0.175 & \textbf{0.211} & 0.194 & 0.201 & 0.187 &  &  \\
\bottomrule

\end{tabular}
\caption{Kruskal–Wallis test results comparing all five teams across metrics ($\rho$, $C_{OD}$, $C_{ID}$). Boldfaced values indicate the highest density and the lowest centralization. I, Q, D, C denote Inform, Question, Directive, Commissive, respectively. The highest value of density and centralization between teams is boldfaced. Values are shown as Mean (SD), with test statistics $H$ and corresponding $p$-values (* $p<.05$, ** $p<.01$, *** $p<.001$).}
\label{tab:kruskal}
\Description{Table 4: Results from the between-team comparison using the Kruskal-Wallis test is shown in Table 4. Density and centralization of all DA pairs showed significant differences across five teams.}
\end{table*}

Results on the Kruskal–Wallis test on density and centralization values between five teams are shown in Table \ref{tab:kruskal}. Between-group comparison revealed significant differences in density across the five teams for all DA pairs. Post hoc comparisons showed that professional team A consistently exhibited higher density than all other teams except professional team C for the Inform → Inform, Question → Inform, Directive → Commissive, and Inform → Directive pairs (p < .01). For Question → Inform and Directive → Commissive pairs, professional team B demonstrated significantly lower density compared to professional team C and amateur team B (p < .01).

With respect to centralization, professional team A consistently recorded the lowest levels of both indegree and outdegree centralization across all DA pairs. Post hoc tests indicated that professional team B showed significantly higher indegree and outdegree centralization than professional team A for the Question → Inform, Directive → Commissive, Inform → Directive, and Inform → Question pairs (all p < .001), whereas no significant difference was observed for the Commissive → Inform pair. Moreover, professional team B exhibited lower centralization than professional team C in indegree centralization for the Question → Inform pair and in both indegree and outdegree centralization for the Directive → Commissive pair, while the two teams did not differ significantly for the Inform → Inform, Inform → Question, and Commissive → Inform pairs. In contrast, professional team B did not differ significantly from amateur team A in either indegree or outdegree centralization across any DA pair. Taken together, these findings suggest that professionally trained teams are generally characterized by higher density and lower centralization, although professional team B more closely resembled amateur teams in showing lower density and higher centralization. Thus, H3 is partially supported.

\begin{table*}
\centering
\begin{tabular}{lccccc|cc}
\hline
 Survey items & Pro A & Pro B & Pro C & Amateur A & Amateur B & $H$ & $p$\\
\hline
Q1: Communication effectiveness & 4.4 (1.14) & 2.8 (1.79) & 3.8 (1.30) & 4.2 (1.64) & 4.67 (1.21) & 3.44  & 0.49 \\
Q2: Individual participation    & 5.4 (1.52) & 4.4 (1.95) & 5.4 (1.14) & 4.8 (1.48) & 4.67 (1.63) & 1.25  & 0.87 \\
Q3: Team participation          & 5.0 (0.71) & 2.8 (0.84) & 5.4 (1.14) & 5.8 (1.10) & \textbf{5.83 (0.75)} & 13.14 & 0.01* \\
Q4: Speaker dominance           & 3.6 (1.14) & \textbf{5.6 (0.89)} & 4.6 (0.55) & 5.2 (1.30) & 4.0 (1.10)  & 10.51 & 0.03* \\
Q5: Communication effectiveness (MoI) & 3.8 (0.84) & 3.0 (1.58) & 4.4 (1.14) & 4.4 (1.14) & 4.5 (1.76) & 3.44  & 0.49 \\
Q6: Team participation (MoI)    & 4.6 (1.14) & 3.8 (1.64) & 5.2 (1.10) & 5.2 (1.92) & 4.67 (1.51) & 2.93  & 0.57 \\
Q7: Speaker dominance (MoI)     & 4.8 (1.48) & 2.6 (1.34) & 4.2 (1.10) & 4.2 (0.84) & 4.0 (1.10)  & 3.12  & 0.54 \\
\hline
\end{tabular}
\caption{Survey results from five teams on teams' perceived communication pattern. 7-Likert values are presented as Mean (SD), with corresponding Kruskal-Wallis $H$ statistics and $p$-values (* $p<.05$).}
\label{tab:survey}
\Description{Table 5: Survey results and statistical values of the Kruskal-Wallis test are shown in Table 5. Significant differences were observed in Q3: Team participation and Q4: Speaker dominance.}
\end{table*}
\section{Discussion}
This study presents a framework that leverages in-game data and voice communication data to automatically model the communication networks of esports teams and to introduce metrics for quantitatively analyzing communication patterns. 
In collaboration with domain experts, we defined MoIs from in-game data and proposed criteria for dividing a game into four distinct phases. 
In addition, the verbal communication analysis module, which segments utterances and classifies DAs from audio data, achieved reasonable accuracy. 
The integration of MoI with SNA offers a systematic approach for analyzing context-dependent communication patterns.
In this section, based on interviews with esports experts, we explain whether the analysis results are consistent with the characteristics typically observed in team communication.
Also, we describe, based on survey responses from players of each team, whether the results align with their own evaluations of their team’s communication effectiveness and distinctive features.

\subsection{Validation of Communication Network Metrics through Esports Expert Insights}
To validate the meaning and applicability of the metrics we proposed (i.e., density, centralization) in this study, we conducted interviews with esports experts based on the analysis results. 
First, the experts noted that the more players participate in the communication process for collaboration, the more likely the team is to achieve good performance. 
This observation aligns with the density-performance hypothesis~\cite{balkundi2006ties}, which argues that greater participation in communication increases interdependence, thereby enabling greater trust and information sharing. 
In addition, centralization refers to the extent to which utterances are concentrated on a particular player. 
The experts mentioned that the centralization can be interpreted as the degree of reliance on that player. 
They further explained that “\textit{as teams advance in skill level, players acquire the ability to make the right utterances at the right timing, and thus the tendency to rely on specific individuals diminishes.}” 
They also noted that “\textit{as teams progress from lower to higher tier teams, all members tend to contribute utterances more evenly in order to make optimal decisions efficiently and effectively.}”
Conversely, teams that rely heavily on a single player’s communication are vulnerable, as the absence of that player may critically undermine team performance. 
Finally, the experts stated that “\textit{in the ideal form of team communication pursued by coaches, every player possesses, and under such conditions all players participate in communication in a balanced manner.}”

In addition, we identified collaborative patterns that frequently appeared in players’ conversations based on audio and in-game data, and confirmed that these patterns corresponded to critical cooperative situations in actual gameplay. 
In particular, experts noted that the two DA pairs (i.e., Question $\rightarrow$ Inform and Directive  $\rightarrow$ Commisive) that showed significant differences between MoI and non-MoI each indicated a unique form of collaboration activity. 
For instance, the Question $\rightarrow$  Inform pair can be interpreted as collaborative information seeking and sharing. 
According to the experts, this pattern represents a key component of problem-solving in situations requiring cooperation, functioning as communication that checks the situation before proposing collaborative actions.
For instance, in a team fight MoI, a dense Question $\rightarrow$ Inform network enables the team to obtain comprehensive information from multiple players, thereby reducing vulnerability.
On the other hand, the Directive $\rightarrow$  Commissive pair can be interpreted as a situation in which a particular player issues directives to teammates, who then respond when making team-level decisions.
Experts further emphasized that directive and response patterns occur frequently before and during cooperative situations. 
Such communication represents a critical type of communication that plays a decisive role in determining whether collaboration succeeds or fails. 
Experts explained that effective collaboration often sequences these patterns: starting with Question $\rightarrow$ Inform for reconnaissance, followed by Directive $\rightarrow$ Commisive for execution.
Moreover, in both situations (i.e., Question $\rightarrow$ Inform and Directive  $\rightarrow$ Commisive), they indicated that communication can be considered more ideal when more players are engaged (i.e., higher density) and when utterances are not overly concentrated on specific individuals (i.e., lower centralization).

Finally, regarding the temporal dynamics of communication, experts identified the 14-minute mark as the most critical turning point when laning phases end and team strategies shift toward objectives like Dragon, Rift Herald, Atakhan, or Baron. 
The pronounced increase in Directive $\rightarrow$ Commissive sequences as games progress suggests that teams prioritize explicit coordination when decision complexity and time pressure mount. 
This shift from information gathering (Question $\rightarrow$ Inform) to action coordination (Directive $\rightarrow$ Commissive) reflects the changing demands of MOBA gameplay, moving from a focus on individual skill to requirements for team coordination. 
Furthermore, in the comparative analysis between professional and amateur teams, only the Directive $\rightarrow$ Commissive pattern showed significant differences in density, outdegree centralization, and indegree centralization in relation to game performance (p < .01, p < .05, and p < .01, respectively). 
However, when comparing communication network metrics between teams, significant differences were observed across all three metrics for all patterns. 
This indicates that professional communication training does not necessarily lead teams to adopt identical communication patterns. 
According to the experts, this variation may stem from differences in the methods of communication training provided and in the extent to which players adhere to such training.

\subsection{Analyzing Team Differences through Player Surveys and Follow-up Interviews}

\begin{figure*}
\centering
  \includegraphics[width=1\linewidth]{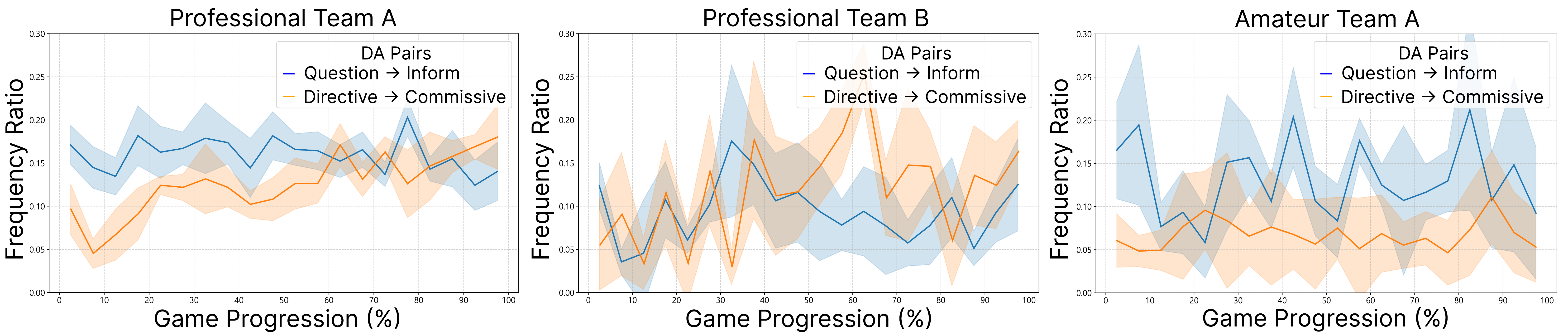}
  \caption{Frequency ratio of Question $\rightarrow$ Inform and Directive $\rightarrow$ Commissive pair across game progression of professional team A, B, and amateur team A. The shaded areas around each line represent the 95\% confidence intervals.}
  \label{fig:discussion_line}
  \Description{Figure 7: Line graph of frequency ratios of Question → Inform and Directive → Commissive pairs along with 95\% confidence interval bands are illustrated in Figure 7. Professional team A and B showed an increase in Directive → Commissive pairs as the game progressed, whereas the frequency of this pair in amateur team A was consistently lower than Question → Inform pair throughout the entire game.}
\end{figure*}

To further analyze why significant differences in communication patterns emerged even among professionally trained teams, we conducted a survey with players from each team using a 7-point Likert scale. 
The survey consisted of seven questions (see Table~\ref{tab:survey}):
\begin{itemize}
    \item \textbf{Q1 - Communication effectiveness:} Our team communicates effectively.
    \item \textbf{Q2 - Individual participation:} I actively participate in team voice chat.
    \item \textbf{Q3 - Team participation:} Our team engages in team voice chat actively.
    \item \textbf{Q4 - Speaker dominance:} One speaker dominates voice chat in our team.
    \item \textbf{Q5 - Communication effectiveness (MoI):} Our team communicates effectively during team fights.
    \item \textbf{Q6 - Team participation (MoI):} Our team engages in team voice chat actively during team fights.
    \item \textbf{Q7 - Speaker dominance (MoI):} One speaker dominates voice chat in our team during team fights.
\end{itemize}
Responses were statistically compared to analyze the group differences using the Kruskal–Wallis test. 
Results revealed a significant difference in the question Q3 asking whether players actively participated in team communication (p < .05). 
Professional team B reported the lowest values, while amateur team B reported the highest. 
To identify specific group differences, we conducted pairwise Mann–Whitney U tests as a post hoc analysis. 
The analysis confirmed that professional team B reported significantly lower values than professional teams A and C, as well as amateur team B, regarding the extent of active participation in communication (p < .05 for all cases). 
This finding aligns with the lower density observed for professional team B compared to the other professional teams.
In addition, significant differences were also found in the question Q4 asking whether communication tended to be dominated by a single player (p < .05). 
Post hoc tests revealed that professional team B reported significantly higher values than professional teams A and C (p < .05), and, overall, recorded the highest value among all teams. 
This result aligns with the higher centralization observed in professional team B’s communication networks compared to professional teams A and C. 
Finally, although no significant between-team differences were found in the question asking players to self-evaluate communication effectiveness, professional team B reported the lowest value (i.e., 2.8 on the 7-point scale).

In follow-up interviews conducted after the survey, professional team B expressed the greatest dissatisfaction with their communication training. 
For example, player P7 noted, “\textit{It was disappointing that the coach provided too little feedback during communication training.}” 
Similarly, players P6, P8, and P9 responded that they felt “\textit{there was a lack of detailed explanation about what went wrong in communication and how it could be improved.}”
Taken together, these findings indicate a clear correspondence between our network-based analysis results and the survey responses collected from players. Moreover, the quality of training is a critical factor in practicing broader and more balanced communication.

Furthermore, as illustrated in Fig.~\ref{fig:discussion_line}, the frequency ratio of the Directive → Commissive pair in professional teams A and B showed a marked increase over the course of the game: it was substantially lower than the Question → Inform pair in the early phase, but surpassed it toward the end of the match. In contrast, amateur teams exhibited consistently lower frequencies of Directive → Commissive pairs compared to Question → Inform pairs throughout the entire game. Experts interpreted these patterns as reflecting the training emphasis placed on professional teams: as games progress and collaborative demands increase, trained teams are encouraged to engage in communication that involves making proposals and confirming them, thereby reaching optimal decisions. Moreover, the confidence intervals further indicate that both professional team B and amateur team A displayed greater variability across games for each pair, relative to professional team A. The relatively narrow confidence intervals for professional team A suggest that its communication patterns were more consistent and systematic across games.

\section{Limitation}
\yc{The method of assessing verbal communication patterns using voice data and in-game data, and the speech recognition model proposed in our research contributes to large-scale quantitative analysis of team-level communication among esports players.
However, several limitations should be noted.
First, the framework that integrates multi-stage modeling may lead to information loss during the process of transcribing and classifying audio data.
Moreover, the fast-paced nature of esports communication often leads to lower accuracy compared to existing benchmarks.
Nevertheless, the findings from RQ2, which aligned with collaborative patterns validated by esports experts demonstrates the applicability of our approach.
Second, to minimize the influence of trust and familiarity associated with team formation duration, we recruited teams that had been formed for less than one month. However, this design choice precluded the possibility of examining how verbal communication patterns evolve as teams mature over time.
Future research could address this limitation by leveraging our logger system and framework to conduct longitudinal investigations into how trust and familiarity shape verbal communication.}

\section{Conclusion}
\yc{In this study, we proposed a framework capable of automatically assessing esports players’ verbal communication patterns in the context of collaboration by utilizing in-game data and voice data.
We conducted a data collection study using the logging system and explored diverse conditions that influence verbal communication patterns in competitive gameplay.
As a result, more players engaged in utterances related to information sharing and action negotiation, and their participation was more evenly distributed in situations requiring collaboration.
Professionally trained teams also demonstrated broader and more balanced contributions to communication; however, this effect was not uniformly observed, suggesting that the effectiveness of training plays a critical role.
Findings from this study can be applied to diagnose team communication effectiveness and inform training practices. Moreover, the framework can also be extended to broader contexts as a generalizable approach for analyzing CMC.}

\bibliographystyle{ACM-Reference-Format}
\bibliography{sample-base}


\appendix









\end{document}